\documentclass[aps,prd,reprint,nofootinbib,superscriptaddress,longbibliography,floatfix,raggedbottom]{revtex4-2}

\usepackage{amsmath,amssymb,amsfonts,bm,mathtools,mathrsfs}
\usepackage{graphicx}
\usepackage{booktabs}
\usepackage{array}
\usepackage{multirow}
\usepackage{hyperref}
\usepackage{microtype}
\usepackage{placeins}

\hypersetup{colorlinks=true,citecolor=blue,urlcolor=blue,linkcolor=blue,
pdftitle={Deviations from Kerr: Critical curves and photon-ring critical parameters for general observer inclinations},
pdfauthor={Hao-Peng Yan; Xiang-Qian Li; Xiao-Jun Yue},
pdfsubject={Analytic deviations of critical curves and photon-ring critical parameters from Kerr},
pdfkeywords={black holes, photon rings, critical curves, separable spacetimes}}
\newcommand{\dd}{\mathrm d}
\newcommand{\calP}{\mathcal P}
\newcommand{\calQ}{\mathcal Q}
\newcommand{\calW}{\mathcal W}
\newcommand{\calH}{\mathcal H}
\newcommand{\calR}{\mathcal R}
\newcommand{\calT}{\mathcal T}
\newcommand{\calV}{\mathcal V}
\newcommand{\K}{\mathrm K}
\newcommand{\ord}{\mathcal O}
\newcommand{\Copt}{\widetilde{\mathscr C}}
\newcommand{\rK}{\tilde r_{\K}}
\newcommand{\Heav}{\mathsf H}
\newcommand{\vect}[1]{\bm{#1}}

\begin{document}

\title{Deviations from Kerr: Critical curves and photon-ring critical parameters for general observer inclinations}

\author{Hao-Peng Yan}
\email[Corresponding author: ]{yanhaopeng@tyut.edu.cn}
\author{Xiang-Qian Li}
\author{Xiao-Jun Yue}
\affiliation{College of Physics and Optoelectronic Engineering, Taiyuan University of Technology, Taiyuan 030024, China}

\begin{abstract}
We study critical curves and photon-ring critical parameters in stationary, axisymmetric spacetimes with separable null geodesics and Kerr angular structure.  A Kerr-relative decomposition of the radial metric functions gives analytic first-order departures of these observables for observers at general inclination.  The critical curve's signed normal displacement factorizes as $\tilde u_n=\tilde F_{\K}\Copt$, where $\tilde F_{\K}$ depends only on the Kerr reference geometry and $\Copt$ is an explicit combination of optical metric deviations.  The demagnification exponent, time delay, and azimuthal advance are evaluated at a common position angle on the observer screen.  Intrinsic relations among the exact observables express the optical information shared by the curve, time delay, and azimuthal advance, while demagnification additionally depends on radial instability.  Applications to representative metrics illustrate how different radial deformations produce distinct departures from Kerr around the critical curve.  Comparisons with exact finite-deformation profiles and independent numerical calculations verify the first-order predictions on regular, accessible exterior photon-shell branches.
\end{abstract}

\maketitle

\section{Introduction}
\label{sec:intro}

Unstable spherical photon orbits connect the strong-field geometry of a rotating black hole to its horizon-scale lensing structure.  In Kerr spacetime, separability organizes these orbits into a one-parameter photon shell \cite{Kerr1963,Carter1968,Teo2003}.  Rays that approach this shell generate increasingly narrow high-order images whose limiting screen locus is the critical curve \cite{Bardeen1973,GrallaHolzWald2019,GrallaLupsasca2020}.  Three critical parameters describe the associated near-critical dynamics: the demagnification exponent $\tilde\gamma$, time delay $\tilde\tau$, and azimuthal advance $\tilde\delta$ per polar half-libration \cite{GrallaLupsasca2020,Johnson2020}.  Their angular dependence carries information about both the location of the trapped rays and the radial rate at which neighboring trajectories escape.

Horizon-scale images of M87* and Sgr~A* provide the observational motivation for testing this geometry \cite{EHTM87,EHTSgrA}.  The critical curve must be distinguished from the emission-dependent brightness depression and from any individual finite-order image \cite{GrallaHolzWald2019,ChaelEtAl2021}.  Nevertheless, narrow photon subrings have a characteristic long-baseline interferometric signature, and their projected diameters can approach those of the critical curve for suitable emission models \cite{Johnson2020,GrallaMeasuring2020,GrallaLupsascaMarrone2020}.  Shape reconstruction and finite-order corrections have been studied analytically and with adaptive ray tracing \cite{GrallaObservable2020,PaugnatEtAl2022,CardenasLupsasca2023,CardenasEtAl2023AART}.  Space-VLBI concepts such as the Black Hole Explorer target these scales \cite{BHEX2024,BHEXScience2024}.  Time-dependent lensing and photon-ring autocorrelations offer complementary access to the orbital time and azimuthal scales \cite{Wong2021,Hadar2021}, although finite ring width and source structure remain relevant to their interpretation \cite{SalehiBroderick2026}.

Parametrized rotating metrics describe departures from Kerr in terms of functions whose optical effects can be studied before choosing a particular physical model \cite{JohannsenPsaltis2010,CardosoPaniRico2014,KonoplyaEtAl2016}.  The radial separable class offers a controlled setting in which the null equations remain analytic \cite{Johannsen2015,JohannsenPhotonRings2013,PapadopoulosKokkotas2018,CarsonYagi2020,YagiEtAl2024}.  A polar observer selects one spherical orbit, whereas finite inclination samples a visible interval of the shell.  General-inclination observables consequently vary around the screen.  At fixed mass and spin, a deformation can alter both the overall size and the relative changes along the prograde and retrograde sides.  Comparing points at equal coordinate radius mixes this geometric change with a change of the shell-to-screen correspondence.  Holding the screen direction fixed, or taking the normal displacement of the curve, separates these effects.

Several results establish the exact starting point for this calculation.  General shadow-boundary formulas for Kerr-like radial mass-function metrics, including rotating Bardeen geometries, were obtained and used to study their size and shape in Refs.~\cite{TsukamotoLiBambi2014,Tsukamoto2018}.  Nonparametric analyses relate polar shadow size and photon-ring instability to radial metric information \cite{BroderickSalehiGeorgiev2023,SalehiBroderickGeorgiev2024}.  Salehi et al. derive general-inclination critical parameters for integrable metrics, and Walia et al. study their dependence on spin, inclination, and explicit non-Kerr families \cite{SalehiWaliaChangKocherlakota2025,WaliaKocherlakotaChangSalehi2025}.  The Kerr-off-shell construction gives exact critical curves with radial and angular deformations \cite{AchourGourgoulhonRoussille2025}.  Perturbative methods address light-ring shifts in post-Kerr spectroscopy and shadow descriptors in a spin expansion \cite{GlampedakisPappasSilvaBerti2017,KobialkoGaltsov2026}.  The question considered here is how general radial departures from Kerr change the unparameterized critical curve and its direction-resolved critical parameters at first order.  This requires separating the physical normal deformation from the displacement produced by matching spherical orbits at the same radial coordinate.

We use the metric conventions and optical radial functions of the companion polar analysis \cite{YanEtAl2026Polar} to derive Kerr-relative changes over the full visible shell interval.  The radial decomposition distinguishes the optical functions that determine the critical curve from the kinetic factor entering radial instability.  At first order, the normal deformation factorizes as $\tilde u_n=\tilde F_{\K}\Copt$, with an explicit metric-deviation function $\Copt$ and a known Kerr geometrical factor.  The accompanying formulas give the radius, demagnification exponent, time delay, and azimuthal advance at a common observer-screen direction.  Intrinsic relations among the exact observables explain how the optical deformation and radial instability enter these changes.  Representative metrics illustrate the resulting angular profiles, including their behavior when the polar critical-circle radius is matched.  Exact finite-deformation calculations and independent numerical checks test the analytic predictions.

Section~\ref{sec:metric} defines the metric functions and null geodesics.  Section~\ref{sec:exact} gives the exact critical observables and the relations between them.  Section~\ref{sec:weak} derives their first-order Kerr-relative changes, and Sec.~\ref{sec:results} applies the formulas to representative metrics.  Section~\ref{sec:discussion} discusses the implications and limits.  The appendices contain radial-relabeling and polar-limit relations and projected-diameter formulas.
\section{Metric representations and null geodesics}
\label{sec:metric}

We consider stationary, axisymmetric, equatorially reflection-symmetric, asymptotically flat spacetimes with separable vacuum null geodesics and Kerr angular structure.  We adopt the radial metric conventions of Ref.~\cite{YanEtAl2026Polar} and define the functions needed for general-inclination observables below.

\subsection{Metric representations}
\label{subsec:opticalmetric}

We work in Boyer--Lindquist-like coordinates $(t,r,\theta,\phi)$ with mass $M$ and spin parameter $a=J/M$.  We choose the azimuthal orientation so that $a>0$; the Schwarzschild limit is treated separately.  We write the radial separable metric using a generalized $\Delta(r)$, Kerr-relative functions $a_i(r)$ ($i=1,2,5$), and the conformal function $f(r)$ as
\begin{align}
\dd s^2={}&
-\frac{\Sigma\left[\Delta-a^2a_2^2\sin^2\theta\right]}
{\left[(r^2+a^2)a_1-a^2a_2\sin^2\theta\right]^2}\,\dd t^2
\nonumber\\
&-\frac{2a\Sigma\sin^2\theta\left[(r^2+a^2)a_1a_2-\Delta\right]}
{\left[(r^2+a^2)a_1-a^2a_2\sin^2\theta\right]^2}\,\dd t\,\dd\phi
\nonumber\\
&+\frac{\Sigma\sin^2\theta\left[(r^2+a^2)^2a_1^2-a^2\Delta\sin^2\theta\right]}
{\left[(r^2+a^2)a_1-a^2a_2\sin^2\theta\right]^2}\,\dd\phi^2
\nonumber\\
&+\frac{\Sigma}{\Delta a_5}\,\dd r^2+\Sigma\,\dd\theta^2,
\label{eq:Johannsenmetric}
\end{align}
where all radial functions are evaluated at $r$ and
\begin{equation}
\Sigma=r^2+a^2\cos^2\theta+f(r),
\qquad
\Delta_{\K}=r^2-2Mr+a^2.
\label{eq:Sigma}
\end{equation}
Separating the Kerr radial function gives
\begin{equation}
\Delta(r)=\Delta_{\K}(r)+h(r).
\label{eq:Deltasplit}
\end{equation}
The Kerr limit is $f=h=0$ with $a_1=a_2=a_5=1$.  This representation makes changes of $\Delta(r)$ and the factors $a_i(r)$ explicit, as needed for the metric examples in Sec.~\ref{sec:results}.

We treat these quantities as general radial functions rather than imposing a particular finite-order parametrized expansion.  Continued-fraction and more general axisymmetric parametrizations provide alternative descriptions of strong-field departures \cite{RezzollaZhidenko2014,KonoplyaEtAl2016}; separability and the undeformed Kerr angular sector are the additional restrictions used here.  They retain spherical photon orbits and closed angular integrals, but exclude generic nonseparable trapping dynamics \cite{GlampedakisPappas2019}.  The metric functions are assumed smooth on the exterior interval of interest, with a regular Lorentzian metric and the usual asymptotic normalization of $t$ and $\phi$.

For the null problem, define the optical functions and radial kinetic factor
\begin{equation}
\calP=\frac{(r^2+a^2)a_1}{\sqrt\Delta},
\qquad
\calQ=\frac{aa_2}{\sqrt\Delta},
\qquad
\calW=\Delta a_5.
\label{eq:lowerPQW}
\end{equation}
The inverse metric is
\begin{align}
\Sigma g^{tt}&=-\calP^2+a^2\sin^2\theta,
&\Sigma g^{t\phi}&=a-\calP\calQ,
\nonumber\\
\Sigma g^{\phi\phi}&=\csc^2\theta-\calQ^2,
&\Sigma g^{rr}&=\calW,
\qquad \Sigma g^{\theta\theta}=1.
\label{eq:inversemetric}
\end{align}
Thus $(\calP,\calQ)$ package the optical information entering the critical impact parameters, whereas $\calW$ fixes the radial kinetic normalization.  This representation is equivalent to the nonparametric radial notation used in Refs.~\cite{SalehiBroderickGeorgiev2024,SalehiWaliaChangKocherlakota2025,WaliaKocherlakotaChangSalehi2025}.\footnote{For their functions $N,F,B_{\rm SBG}$, the dictionary is $\calP=r/N$, $\calQ=aF/N$, and $\calW=r^2N^2/B_{\rm SBG}^2$.  Their non-negative Carter quantity obeys $\mathscr I^2=\eta+(\xi-a)^2=\zeta^2$.  The exact double-root and critical-parameter formulas therefore agree after matching these conventions.}  The Kerr values are
\begin{equation}
\calP_{\K}=\frac{r^2+a^2}{\sqrt{\Delta_{\K}}},
\qquad
\calQ_{\K}=\frac{a}{\sqrt{\Delta_{\K}}},
\qquad
\calW_{\K}=\Delta_{\K}.
\label{eq:kerrPQW}
\end{equation}

On a common exterior interval with $\Delta>0$ and $\Delta_{\K}>0$, the equivalent Johannsen radial functions \cite{Johannsen2015}, with the angular functions fixed to Kerr, are
\begin{equation}
A_1=a_1\sqrt{\frac{\Delta_{\K}}{\Delta}},
\qquad
A_2=a_2\sqrt{\frac{\Delta_{\K}}{\Delta}},
\qquad
A_5=a_5\frac{\Delta}{\Delta_{\K}}.
\label{eq:Kerrrelativemap}
\end{equation}
The optical functions therefore admit the exact Kerr factorizations
\begin{equation}
\calP=\calP_{\K}A_1,
\qquad
\calQ=\calQ_{\K}A_2,
\qquad
\calW=\calW_{\K}A_5.
\label{eq:JohannsenPQW}
\end{equation}
The $a_i$ representation keeps changes of $\Delta$ explicit, while the $A_i$ give the full multiplicative changes of the optical functions and radial kinetic factor.  These relations hold before the weak-deviation expansion.

We adopt the radial-coordinate and asymptotic mass--spin conventions of Ref.~\cite{YanEtAl2026Polar} and hold the ADM mass $M$ and angular momentum $J=Ma$ fixed in all comparisons.  In an asymptotically spherical chart with normalized $t$ and $\phi$, we require
\begin{equation}
\begin{aligned}
g_{tt}&=-1+\frac{2M}{r}+\ord(r^{-2}),\\
g_{t\phi}&=-\frac{2Ma\sin^2\theta}{r}+\ord(r^{-2}),
\end{aligned}
\label{eq:asymptoticcharges}
\end{equation}
together with spatial asymptotic flatness and the falloff needed for well-defined ADM charges.  The limits $\Delta/\Delta_{\K}\to1$, $a_i\to1$, and $f/r^2\to0$ alone do not fix these charges; the leading coefficients of the full metric must be matched.  Changing a mass or spin coefficient changes the Kerr reference and is excluded from the fixed-$(M,a)$ comparisons below.
\subsection{Separated null geodesics}
\label{subsec:geodesics}

Using the Hamilton--Jacobi ansatz
\begin{equation}
S=-Et+L_z\phi+S_r(r)+S_\theta(\theta),
\end{equation}
we define the energy-normalized impact parameters
\begin{equation}
\xi=\frac{L_z}{E},
\qquad
\eta=\frac{\mathcal K}{E^2},
\qquad
\zeta^2=\eta+(\xi-a)^2,
\label{eq:impact}
\end{equation}
and the radial optical combination
\begin{equation}
\calH(r;\xi)=\calP(r)-\xi\calQ(r).
\label{eq:Hdef}
\end{equation}
With $\dd\nu=E\,\dd\lambda/\Sigma$, where $\lambda$ is an affine parameter, the null geodesic equations take the compact first-order form
\begin{align}
\frac{\dd t}{\dd\nu}&=\calT,
&
\frac{\dd r}{\dd\nu}&=\pm\sqrt{\calR},
\nonumber\\
\frac{\dd\theta}{\dd\nu}&=\pm\sqrt{\Theta},
&
\frac{\dd\phi}{\dd\nu}&=\Phi.
\label{eq:geodesicscompact}
\end{align}
The signs in the radial and polar equations select the corresponding trajectory branches and reverse at turning points.  The four separated functions are
\begin{subequations}
\label{eq:geodesicfunctions}
\begin{align}
\calT(r,\theta;\xi)
&=\calP(r)\calH(r;\xi)
+a\bigl(\xi-a\sin^2\theta\bigr),
\label{eq:Tfunc}\\
\calR(r;\xi,\eta)
&=\calW(r)\left[\calH^2(r;\xi)-\zeta^2\right],
\label{eq:Rpot}\\
\Theta(\theta;\xi,\eta)
&=\eta+a^2\cos^2\theta-\xi^2\cot^2\theta,
\label{eq:Thetapot}\\
\Phi(r,\theta;\xi)
&=\calQ(r)\calH(r;\xi)
+\xi\csc^2\theta-a.
\label{eq:Phifunc}
\end{align}
\end{subequations}
The critical-curve location depends on $\calP$ and $\calQ$, whereas $\calW$ enters the radial instability.  A finite nonzero $\calW$ therefore does not move a spherical double root, but it does affect the rate at which nearby rays depart from it.  The functions $\calT$ and $\Phi$ will also enter directly in the time and azimuth accumulated over a polar half-libration.  The conformal factor $\Sigma$ cancels from the null Hamilton--Jacobi equation and from these coordinate trajectories after the change to $\nu$.  Consequently, a regular change of $f(r)$ alone does not affect the vacuum critical observables at fixed asymptotic normalization.  This statement concerns the null trajectories, not timelike motion or the emission physics of a source.

We use geometrized units $G=c=1$.  Thus $\xi$, $\calP$, and $\calH$ have dimensions of length, $\eta$ and $\calW$ of length squared, and $\calQ$ is dimensionless.  The impact parameters are energy-normalized, not dimensionless; numerical profiles below are expressed in units of $M$.
\section{Photon shell and critical observables}
\label{sec:exact}

Double roots of the separated radial potential determine the spherical photon shell and its critical curve.  The radial curvature and the time and azimuth accumulated over a polar half-libration supply the three critical parameters.  Eliminating the radial shell label then exposes the optical information common to these observables.

\subsection{Spherical photon shell and critical curve}
\label{subsec:shellcurve}

A tilde denotes a spherical critical orbit or a function evaluated on it.  For a radial function $F(r)$ we write
\begin{equation}
\tilde F^{(n)}(\tilde r)
:=\left.\frac{\dd^n F}{\dd r^n}\right|_{r=\tilde r},
\qquad n=0,1,2,\ldots,
\end{equation}
with $\tilde F=\tilde F^{(0)}$, $\tilde F'=\tilde F^{(1)}$, and $\tilde F''=\tilde F^{(2)}$.  Radial and impact-parameter derivatives are taken before evaluation.  The same convention applies to functions of several variables: $\widetilde{\Theta}(\theta)=\Theta(\theta;\tilde\xi,\tilde\eta)$, for example, retains the angular argument while fixing the critical impact parameters.  The Kerr subscript specifies the reference geometry independently of the tilde.

A spherical photon orbit at $r=\tilde r$ satisfies
\begin{equation}
\tilde{\calR}=0,
\qquad
\widetilde{\partial_r\calR}=0.
\label{eq:doubleroot}
\end{equation}
On the regular exterior branch with $\tilde{\calW}\neq0$, $\tilde{\calH}>0$, and $\tilde{\calQ}'\neq0$, these conditions reduce to
\begin{equation}
\tilde\xi(\tilde r)=\frac{\tilde{\calP}'}{\tilde{\calQ}'},
\label{eq:xiexact}
\end{equation}
\begin{equation}
\tilde\eta(\tilde r)
=\tilde{\calH}^{\,2}-(\tilde\xi-a)^2,
\qquad
\tilde{\calH}=\tilde{\calP}-\tilde\xi\tilde{\calQ}.
\label{eq:etaexact}
\end{equation}
To see why $\calW$ drops out, first impose $\tilde{\calR}=0$, which gives $\tilde{\calH}^{\,2}=\tilde\zeta^{\,2}$.  The derivative of $\calW$ then multiplies a vanishing bracket in $\widetilde{\partial_r\calR}$.  The remaining condition is $2\tilde{\calW}\tilde{\calH}(\tilde{\calP}'-\tilde\xi\tilde{\calQ}')=0$, yielding Eq.~\eqref{eq:xiexact} on the stated branch.  The critical impact parameters are therefore determined before the radial instability is calculated.

For an asymptotic observer at inclination $0<\theta_o\leq\pi/2$, the corresponding screen coordinates are
\begin{equation}
\tilde\alpha=-\tilde\xi\csc\theta_o,
\qquad
\tilde\beta_{\pm}=\pm\sqrt{\widetilde B},
\label{eq:screen}
\end{equation}
where
\begin{equation}
\widetilde B
=\tilde\eta+a^2\cos^2\theta_o
-\tilde\xi^{\,2}\cot^2\theta_o.
\label{eq:Bscreen}
\end{equation}
Let $\mathcal S_{\rm ext}$ denote the selected exterior unstable spherical-photon branch accessible from infinity.  In addition to the local double-root conditions, accessibility requires that the critical constants admit an exterior radial path to infinity without an intervening forbidden interval.  For the separated radial equation this entails $\calR(r;\tilde\xi,\tilde\eta)\geq0$ on that path.  We follow the branch continuously from Kerr and assume no change of the accessible trapped set over the compared parameter range.  Its visible subset is
\begin{equation}
\mathcal I_o=\{\tilde r\in\mathcal S_{\rm ext}:\widetilde B(\tilde r)\geq0\}.
\label{eq:visibleinterval}
\end{equation}
The two reflected branches of Eq.~\eqref{eq:screen}, restricted to $\mathcal I_o$, form the complete critical curve of the selected shell.  Multiple accessible shells, throat orbits, or a branch transition require an additional global analysis.  Where the curve is star-shaped with respect to the chosen screen origin, we also use
\begin{equation}
\tilde\rho=\sqrt{\tilde\alpha^2+\tilde\beta^2},
\qquad
\tilde\varphi=\arctan\!\left(\frac{\tilde\beta}{\tilde\alpha}\right).
\label{eq:screenpolar}
\end{equation}
The branch of $\arctan$ is chosen continuously along the screen curve; on the upper branch we take $0<\tilde\varphi<\pi$, with $\tilde\varphi=\pi/2$ at $\tilde\alpha=0$.  The complete Cartesian curve remains the primary geometric object; $\tilde\rho(\varphi)$ is a convenient local representation.

\subsection{Photon-ring critical parameters}
\label{subsec:criticalfields}

We normalize all critical parameters per one polar half-libration, i.e. from one polar turning point to the adjacent turning point, following the Kerr lensing conventions of Ref.~\cite{GrallaLupsasca2020}.  Since the angular potential is unchanged, its integral structure is the same as in Kerr \cite{GrallaNull2020}; the metric dependence enters through the critical impact parameters.  For general impact parameters $(\xi,\eta)$, define the angular integrals
\begin{align}
G_\theta&=\int_{\theta_-}^{\theta_+}\frac{\dd\theta}{\sqrt{\Theta}},
\nonumber\\
G_t&=\int_{\theta_-}^{\theta_+}\frac{\cos^2\theta\,\dd\theta}{\sqrt{\Theta}},
\nonumber\\
G_\phi&=\int_{\theta_-}^{\theta_+}\frac{\csc^2\theta\,\dd\theta}{\sqrt{\Theta}}.
\label{eq:Gdefs}
\end{align}
Writing $u=\cos^2\theta$, the two roots of the polar potential are
\begin{equation}
u_\pm
=\frac{a^2-\eta-\xi^2
\pm\sqrt{(a^2-\eta-\xi^2)^2+4a^2\eta}}
{2a^2}.
\label{eq:upm}
\end{equation}
We use the Kerr-connected, equator-crossing polar-libration branch, with $u_-<0$ and $0<u_+\leq1$; equatorial limits are taken continuously.  With elliptic parameter $m=u_+/u_-$, the half-libration integrals are
\begin{align}
G_\theta&=\frac{2}{\sqrt{-a^2u_-}}\,\mathrm K(m),
\nonumber\\
G_t&=\frac{2u_+}{\sqrt{-a^2u_-}}
\frac{\mathrm K(m)-\mathrm E(m)}{m},
\nonumber\\
G_\phi&=\frac{2}{\sqrt{-a^2u_-}}\,\Pi(u_+|m).
\label{eq:ellipticG}
\end{align}
These expressions are continued analytically at degenerate roots; near pole crossing, the finite composite product $\tilde\xi \tilde G_\phi$ is taken before the limit.  On the shell, $\tilde G_j=G_j(a,\tilde\xi,\tilde\eta)$ for $j\in\{\theta,t,\phi\}$; the roots, turning angles, and elliptic parameter evaluated there are denoted by $\tilde u_\pm$, $\tilde\theta_\pm$, and $\tilde m$.  Define
\begin{equation}
\tilde{\calH}_2
=\tilde{\calP}''-\tilde\xi\,\tilde{\calQ}''.
\label{eq:H2}
\end{equation}
Here the two radial derivatives are taken at fixed impact parameters.  In particular, $\tilde{\calH}_2$ is not the second total derivative of $\tilde{\calH}$ along the shell, where $\tilde\xi$ also varies.  At a double root,
\begin{equation}
\frac12\widetilde{\partial_r^2\calR}
=\tilde{\calW}\tilde{\calH}\tilde{\calH}_2.
\label{eq:radialcurvature}
\end{equation}
A small radial offset consequently grows locally as $\exp[\sqrt{\widetilde{\partial_r^2\calR}/2}\,\nu]$.  Multiplying its growth rate by the angular interval $\Delta\nu=\tilde G_\theta$ gives the dimensionless exponent per half-libration.  This normalization distinguishes the photon-ring exponent from a Lyapunov rate per coordinate time \cite{CardosoEtAl2009,DeichEtAl2024}.  We restrict to unstable exterior orbits with $\tilde{\calW}\tilde{\calH}\tilde{\calH}_2>0$.  The exact direction-resolved critical parameters are
\begin{subequations}
\label{eq:exactfields}
\begin{align}
\tilde\gamma&=\tilde G_\theta
\sqrt{\tilde{\calW}\tilde{\calH}\tilde{\calH}_2},
\label{eq:gammaexact}\\
\tilde\tau&=
\left[\tilde{\calP}\tilde{\calH}+a\tilde\xi-a^2\right]\tilde G_\theta
+a^2\tilde G_t,
\label{eq:tauexact}\\
\tilde\delta&=
\left[\tilde{\calQ}\tilde{\calH}-a\right]\tilde G_\theta
+\tilde\xi \tilde G_\phi
+2\pi\Heav(\tilde r-\tilde r_0).
\label{eq:deltaexact}
\end{align}
\end{subequations}
The time and azimuthal expressions follow by integrating $\calT$ and $\Phi$ between consecutive polar turning points with $r=\tilde r$.  Their radial terms are constant on that orbit, leaving only the three angular integrals above.  These general-inclination critical parameters have been studied in the integrable non-Kerr setting \cite{SalehiWaliaChangKocherlakota2025,WaliaKocherlakotaChangSalehi2025}; the present notation keeps their optical and kinetic ingredients explicit.

The last term in Eq.~\eqref{eq:deltaexact} specifies a continuously unwrapped branch across the pole-crossing orbit $\tilde r_0$ where $\tilde\xi=0$.  The product $\tilde\xi\tilde G_\phi$ has finite one-sided limits there, but its angular-coordinate branch changes across the pole.  Continuity refers to the full unwrapped $\tilde\delta$, including the branch correction, rather than to the isolated product.  Differences and derivatives must use the same unwrapped convention.  The units provide a further check: $\tilde G_j$ has dimensions of inverse length, $\tilde\tau$ of length, and $\tilde\gamma$ and $\tilde\delta$ are dimensionless.

The critical parameters characterize the asymptotic demagnification, time delay, and azimuthal advance associated with additional polar half-librations \cite{GrallaLupsasca2020,Johnson2020}.  Relating them to a particular high-order image sequence requires specifying the source geometry and image labeling.
\subsection{Intrinsic relations among the exact critical observables}
\label{sec:intrinsic}

The critical curve and photon-ring parameters are determined by the same family of spherical orbits.  Their relations can be expressed independently of the chosen radial label by using the conserved impact parameter $\tilde\xi$ as a local shell coordinate.  These are the intrinsic relations considered below.  We work on a regular unstable patch with $\tilde{\calQ}'\neq0$ and $\dd\tilde\xi/\dd\tilde r\neq0$; more general shells can be covered patchwise.

\paragraph{Optical-shell information.}
The spherical condition gives
\begin{equation}
\tilde{\calP}'-\tilde\xi\,\tilde{\calQ}'=0.
\label{eq:sphericalidentity}
\end{equation}
Along the spherical family,
\begin{equation}
\frac{\dd\tilde{\calH}}{\dd\tilde r}
=\tilde{\calP}'-\tilde\xi\,\tilde{\calQ}'-\tilde{\calQ}\frac{\dd\tilde\xi}{\dd\tilde r}
=-\tilde{\calQ}\frac{\dd\tilde\xi}{\dd\tilde r},
\end{equation}
so that
\begin{equation}
\frac{\dd\tilde{\calH}}{\dd\tilde\xi}=-\tilde{\calQ}.
\label{eq:Hxi}
\end{equation}
On the regular positive exterior branch, Eq.~\eqref{eq:etaexact} gives
\begin{equation}
\tilde{\calH}(\tilde\xi)
=\sqrt{\tilde\eta(\tilde\xi)+(\tilde\xi-a)^2},
\label{eq:Hfrometa}
\end{equation}
and therefore
\begin{equation}
\tilde{\calQ}(\tilde\xi)=-\tilde{\calH}_{,\tilde\xi},
\qquad
\tilde{\calP}(\tilde\xi)
=\tilde{\calH}-\tilde\xi\,\tilde{\calH}_{,\tilde\xi}.
\label{eq:PQfrometa}
\end{equation}
Thus the differentiable intrinsic critical locus $\tilde\eta(\tilde\xi)$ fixes the values of the two optical functions sampled along the visible spherical-photon shell, without fixing their off-shell continuation or a preferred radial coordinate.  For specified $(a,\theta_o)$, the same locus follows directly from the complete observer-screen critical curve through
\begin{equation}
\tilde\xi=-\tilde\alpha\sin\theta_o,
\qquad
\tilde\eta
=\tilde\beta^2+(\tilde\alpha^2-a^2)\cos^2\theta_o.
\label{eq:screeninverse}
\end{equation}

\paragraph{Time delay and azimuthal advance.}
Differentiating Eq.~\eqref{eq:Hfrometa} gives
\begin{equation}
\tilde{\calH}\tilde{\calH}_{,\tilde\xi}
=\frac12\tilde\eta_{,\tilde\xi}+\tilde\xi-a.
\label{eq:Hetaidentity}
\end{equation}
Together with Eq.~\eqref{eq:PQfrometa}, this yields
\begin{equation}
\tilde\tau(\tilde\xi)=
\left(\tilde\eta-\frac{\tilde\xi}{2}\tilde\eta_{,\tilde\xi}\right)\tilde G_\theta
+a^2\tilde G_t,
\label{eq:taucurve}
\end{equation}
\begin{equation}
\tilde\delta(\tilde\xi)=
\left(-\tilde\xi-\frac12\tilde\eta_{,\tilde\xi}\right)\tilde G_\theta
+\tilde\xi \tilde G_\phi+2\pi \tilde N_\delta(\tilde\xi),
\label{eq:deltacurve}
\end{equation}
where $\tilde N_\delta$ is chosen patchwise so that the unwrapped $\tilde\delta$ is continuous across the pole-crossing orbit.  Since the angular integrals depend only on $(a,\tilde\xi,\tilde\eta)$ in the present radial class, the complete optical-shell relation fixes the ideal $\tilde\tau$ and $\tilde\delta$ profiles for fixed global parameters.

\paragraph{Additional radial-instability information.}
Differentiating Eq.~\eqref{eq:sphericalidentity} along the spherical family gives
\begin{equation}
\tilde{\calH}_2
=\tilde{\calP}''-\tilde\xi\,\tilde{\calQ}''
=\frac{\dd\tilde\xi}{\dd\tilde r}\tilde{\calQ}'
=-\tilde{\calH}_{,\tilde\xi\tilde\xi}
\left(\frac{\dd\tilde\xi}{\dd\tilde r}\right)^2.
\label{eq:H2intrinsic}
\end{equation}
It is therefore natural to define
\begin{equation}
\tilde{\calV}(\tilde\xi)
=\tilde{\calW}(\tilde\xi)
\left(\frac{\dd\tilde\xi}{\dd\tilde r}\right)^2,
\label{eq:Vdef}
\end{equation}
so that
\begin{equation}
\tilde\gamma(\tilde\xi)
=\tilde G_\theta
\sqrt{-\tilde{\calV}\tilde{\calH}\tilde{\calH}_{,\tilde\xi\tilde\xi}}.
\label{eq:gammaintrinsic}
\end{equation}
Under a smooth radial reparameterization $\bar r=\bar r(r)$, the shell labels satisfy $\bar{\tilde r}=\bar r(\tilde r)$, and the factors $\tilde{\calW}$ and $(\dd\tilde\xi/\dd\tilde r)^2$ transform inversely, leaving $\tilde{\calV}$ invariant.  The exact shell observables considered here are therefore organized by two intrinsic functions,
\begin{equation}
\tilde\eta(\tilde\xi),
\qquad
\tilde{\calV}(\tilde\xi),
\label{eq:minimalintrinsic}
\end{equation}
where the first determines the curve, $\tilde\tau$, and $\tilde\delta$, while the second supplies the additional information entering $\tilde\gamma$.  These relations require a differentiable curve in calibrated screen coordinates and specified $(M,a,\theta_o)$.  Recovering the optical relation from observations requires the full curve \cite{GrallaObservable2020} and known global parameters \cite{AchourGourgoulhonRoussille2025}.  The instability parameter then supplies $\tilde{\calV}$ once the optical locus and its derivatives are specified, without uniquely fixing the off-shell metric.

\paragraph{Strict polar limit.}
The discussion above assumes a finite visible interval on which $\tilde\eta(\tilde\xi)$ is differentiable.  A strict polar observer selects a single pole-crossing spherical orbit with $\tilde\xi=0$, so the observed circle supplies local data at one orbit rather than a function over an interval.  The polar weak-deviation formulas are therefore treated separately in Appendix~\ref{app:relabelpolar}.

\section{Weak deviations from Kerr}
\label{sec:weak}

The exact expressions give the observables once the radial functions of a metric are supplied.  Their first-order expansion isolates how a departure from Kerr changes the curve shape and the critical-parameter profiles.  We hold the physical reference $(M,a,\theta_o)$ fixed and expand only in the non-Kerr deformation parameter; the Kerr spin dependence is retained in the coefficients.  The perturbed branch must remain exterior, accessible, and unstable.  Smallness of a parameter at infinity alone does not ensure a controlled expansion near a degenerate photon orbit.

We write
\begin{align}
\calP&=\calP_{\K}(1+\varepsilon U_1),
&\calQ&=\calQ_{\K}(1+\varepsilon U_2),
\nonumber\\
\calW&=\calW_{\K}(1+\varepsilon U_5),
\label{eq:weakPQW}
\end{align}
where $\varepsilon$ is a bookkeeping parameter and $U_i(r)$ are smooth dimensionless Kerr-deviation profiles on the relevant strong-field interval.  A generic quantity is expanded as
\begin{equation}
X=X_{\K}+\varepsilon X^{[1]}+\ord(\varepsilon^2).
\label{eq:genericweak}
\end{equation}
For the Kerr-relative representation of Sec.~\ref{subsec:opticalmetric},
\begin{equation}
\Delta=\Delta_{\K}(1+\varepsilon\epsilon_\Delta),
\qquad
a_i=1+\varepsilon\epsilon_i,
\label{eq:lowereps}
\end{equation}
so Eq.~\eqref{eq:lowerPQW} gives
\begin{equation}
U_1=\epsilon_1-\frac12\epsilon_\Delta,
\qquad
U_2=\epsilon_2-\frac12\epsilon_\Delta,
\qquad
U_5=\epsilon_5+\epsilon_\Delta.
\label{eq:lowerUmap}
\end{equation}
This decomposition is tied to the selected radial chart and is a calculational dictionary rather than an invariant separation into independently observable metric directions.

For direct application, the metric functions first give $(U_1,U_2,U_5)$ through Eq.~\eqref{eq:lowerUmap}.  Section~\ref{subsec:weakcurve} gives the curve's normal displacement and its radius at a common screen angle.  Section~\ref{subsec:weakfields} computes the other three changes in the same screen convention.  Section~\ref{subsec:weakintrinsic} then rewrites these results in terms of the optical deformation function and radial instability.  Fixed-radius variations are intermediate quantities, whereas fixed-screen variations are the profiles plotted in Sec.~\ref{sec:results}.

\subsection{Kerr-relative critical-curve deviations}
\label{subsec:weakcurve}

For first-order comparisons, $\rK$ denotes a spherical orbit of the Kerr reference, while $\tilde r$ remains the orbit radius in a general metric.  The restricted profiles $\tilde U_i=U_i(\rK)$ and all Kerr quantities below are evaluated at this reference orbit.  A prime denotes a radial derivative evaluated there; $\tilde{\calH}_{\K}'$ and $\tilde\xi_{\K}'$ are derivatives along the Kerr shell.  Derivatives with respect to the impact parameter are written explicitly as subscripts $,\tilde\xi$.

The physically relevant change of the critical curve is the deformation of the \emph{unparameterized} screen curve, rather than the displacement of points carrying the same spherical-orbit label.  We therefore begin with its signed outward normal displacement,
\begin{equation}
\tilde u_n:=\hat{\tilde{\vect n}}_{\K}\cdot\delta\tilde{\vect x},
\label{eq:dmu}
\end{equation}
where $\hat{\tilde{\vect n}}_{\K}$ is the outward unit normal to the Kerr critical curve.  Thus $\tilde u_n>0$ denotes an outward deformation and $\tilde u_n<0$ an inward one; tangential displacements only relabel the curve.

The optical metric deviations enter through the combination
\begin{equation}
\Copt(\rK)
\equiv
\tilde{\calP}_{\K}(\rK)\tilde U_1(\rK)
-\tilde\xi_{\K}(\rK)\tilde{\calQ}_{\K}(\rK)\tilde U_2(\rK).
\label{eq:Copt}
\end{equation}
To obtain its contribution to the curve, use the Kerr spherical-orbit radius as the common intermediate label.  Expanding $\tilde{\calH}=\tilde{\calP}-\tilde\xi\tilde{\calQ}$ at fixed reference radius gives
\begin{equation}
\tilde{\calH}^{[1]}_{\rK}
=\Copt-\tilde{\calQ}_{\K}\tilde\xi^{[1]}_{\rK}.
\label{eq:H1r}
\end{equation}
The two terms in Eq.~\eqref{eq:H1r} separate the explicit optical change produced by the metric deformation from the change induced by the shift of the critical impact parameter.  Although $\Copt$ contains no radial derivative of the perturbation profiles, the label-matched shift of $\tilde\xi$ does.  The cancellation below shows why those derivative terms change the point correspondence along the reference curve without introducing a second independent first-order shape profile.

The spherical condition gives the label-matched shell variations
\begin{align}
\tilde\xi^{[1]}_{\rK}
&=
\frac{(\tilde{\calP}_{\K}\tilde U_1)'
-\tilde\xi_{\K}(\tilde{\calQ}_{\K}\tilde U_2)'}{\tilde{\calQ}_{\K}'},
\nonumber\\
\tilde\eta^{[1]}_{\rK}
&=2\tilde{\calH}_{\K}\tilde{\calH}^{[1]}_{\rK}
-2(\tilde\xi_{\K}-a)\tilde\xi^{[1]}_{\rK}.
\label{eq:labelmatchedvariations}
\end{align}
Here the subscript $\rK$ denotes comparison at the same Kerr shell label, with all radial functions evaluated at $r=\rK$.

The impact-parameter contribution is tangential to the Kerr critical curve.  Using the screen map together with the Kerr shell relation $\tilde{\calH}_{\K}'=-\tilde{\calQ}_{\K}\tilde\xi_{\K}'$, the label-matched screen variation can be arranged as
\begin{align}
\tilde\alpha^{[1]}_{\rK}
&=\frac{\tilde\xi^{[1]}_{\rK}}{\tilde\xi_{\K}'}\,\tilde\alpha_{\K}',
\nonumber\\
\tilde\beta^{[1]}_{\rK}
-\frac{\tilde{\calH}_{\K}}{\tilde\beta_{\K}}\Copt
&=\frac{\tilde\xi^{[1]}_{\rK}}{\tilde\xi_{\K}'}\,\tilde\beta_{\K}',
\label{eq:tangentseparation}
\end{align}
valid on a regular patch with $\tilde\xi_{\K}'\neq0$.  The terms proportional to $\tilde\xi^{[1]}_{\rK}$ therefore point along the Kerr critical-curve tangent and vanish under projection onto the outward normal.  The remaining normal deformation factorizes as
\begin{equation}
\tilde u_n(\rK)=\tilde F_{\K}(\rK;\theta_o)\,\Copt(\rK),
\label{eq:un}
\end{equation}
with
\begin{equation}
\tilde F_{\K}=\frac{\tilde{\calH}_{\K}\csc\theta_o}{
\sqrt{\tilde\beta_{\K}^{\,2}\csc^2\theta_o+
\bigl(\tilde{\calH}_{\K}\tilde{\calQ}_{\K}+
\tilde\xi_{\K}\csc^2\theta_o-a\bigr)^2}}.
\label{eq:Kerrnormalfactor}
\end{equation}
Here $\tilde\beta_{\K}$ denotes the positive upper-branch square root; reflection gives the same positive factor at the corresponding lower-branch point.  On a regular exterior Kerr shell $\tilde{\calH}_{\K}>0$, so $\tilde F_{\K}>0$ and the sign of the normal deformation is fixed directly by $\Copt$.

The independent comparison angle $\varphi$ labels points on the critical curve; it is distinct from a general geodesic's azimuth $\phi$.  For direction-resolved comparisons, let $\tilde r_{\K}(\varphi)$ denote the local inverse of the Kerr screen map, $\tilde\varphi_{\K}[\tilde r_{\K}(\varphi)]=\varphi$.  A scalar shell field $\tilde X$ evaluated at a common observer-screen angle has the first-order deviation
\begin{equation}
\tilde X^{[1]}_{\varphi}(\varphi)
=\left[
\tilde X^{[1]}_{\rK}
-\frac{\dd\tilde X_{\K}/\dd\rK}
{\dd\tilde\varphi_{\K}/\dd\rK}
\,\tilde\varphi^{[1]}_{\rK}
\right]_{\rK=\tilde r_{\K}(\varphi)},
\label{eq:fixedscreen}
\end{equation}
valid on a regular patch with $\dd\tilde\varphi_{\K}/\dd\rK\neq0$.  For the critical-curve radius this reduces to
\begin{equation}
\tilde\rho^{[1]}_{\varphi}(\varphi)
=\left[
\frac{\tilde{\calH}_{\K}\tilde\rho_{\K}}
{\tilde\beta_{\K}^2
+\tilde\xi_{\K}[\tilde{\calH}_{\K}\tilde{\calQ}_{\K}+\tilde D_{\theta,\K}]}
\,\Copt
\right]_{\rK=\tilde r_{\K}(\varphi)},
\label{eq:rho1phi}
\end{equation}
where $\tilde D_{\theta,\K}=\tilde\xi_{\K}\csc^2\theta_o-a$.  The angular shift follows from the screen coordinates,
\begin{equation}
\tilde\varphi^{[1]}_{\rK}
=\frac{\tilde\alpha_{\K}\tilde\beta^{[1]}_{\rK}
-\tilde\beta_{\K}\tilde\alpha^{[1]}_{\rK}}
{\tilde\rho_{\K}^{\,2}},
\label{eq:varphi1r}
\end{equation}
where the required Cartesian variations follow from Eq.~\eqref{eq:tangentseparation}.  Equations~\eqref{eq:un} and \eqref{eq:rho1phi} therefore provide complementary parameterization-independent and observer-angle descriptions of the same first-order shape change.  The subtraction in Eq.~\eqref{eq:fixedscreen} follows by expanding the inverse screen map: the orbit seen at a fixed $\varphi$ shifts by $-\tilde\varphi^{[1]}_{\rK}/\tilde\varphi_{\K}'$ in the reference radial label.  Without this term, two profiles would be compared at the same coordinate radius rather than at the same observed direction.  The normal formulation remains meaningful independently of a star-shaped screen representation, while the fixed-$\varphi$ formulas are applied only where its inverse is regular.

The functional optical null condition $\Copt\equiv0$ describes a locally radially relabeled optical shell on a regular patch with $\tilde{\calQ}'_{\K}\ne0$:
\begin{equation}
\tilde{\calP}^{[1]}=\tilde s\tilde{\calP}'_{\K},\qquad
\tilde{\calQ}^{[1]}=\tilde s\tilde{\calQ}'_{\K}.
\label{eq:nullmap}
\end{equation}
Here $\tilde s=s(\rK)$.  Equation~\eqref{eq:nullmap} is the optical tangent generated by an infinitesimal radial relabeling.  Conversely, $\Copt\equiv0$ on a regular shell patch implies the same local functional relation, whereas an isolated zero of $\Copt$ does not.  The corresponding transformation of $\tilde{\calW}$ is given in Appendix~\ref{app:relabelpolar}.

The normal displacement $\tilde u_n=\tilde F_{\K}\Copt$ in Eq.~\eqref{eq:un} also gives first-order projected positions and diameters by evaluation at the Kerr support points.  Appendix~\ref{app:projection} states this connection and its geometric assumptions.

\subsection{Kerr-relative photon-ring critical parameters}
\label{subsec:weakfields}

At a common Kerr shell label, the critical-parameter shifts follow from variations of the radial coefficients and the angular integrals.  Since the latter depend on $(\tilde\xi,\tilde\eta)$,
\begin{equation}
\tilde G_j^{[1]}
=(\partial_{\tilde\xi}\tilde G_j)_{\K}\tilde\xi_{\rK}^{[1]}
+(\partial_{\tilde\eta}\tilde G_j)_{\K}\tilde\eta_{\rK}^{[1]},
\qquad
j\in\{\theta,t,\phi\}.
\label{eq:G1}
\end{equation}
Here $\tilde G_j^{[1]}$ denotes the variation at fixed Kerr shell label; partial derivatives are evaluated before imposing the critical-orbit values.  The required partial derivatives are evaluated analytically from Eq.~\eqref{eq:ellipticG}.  For $q\in\{\xi,\eta\}$ we differentiate the roots $u_\pm(q)$ and use
\begin{equation}
\frac{\partial\mathrm K}{\partial m}
=\frac{\mathrm E-(1-m)\mathrm K}{2m(1-m)},
\qquad
\frac{\partial\mathrm E}{\partial m}
=\frac{\mathrm E-\mathrm K}{2m},
\label{eq:ellipticderivs}
\end{equation}
together with the derivatives of the finite combination $\tilde J_\phi=\tilde\xi\tilde G_\phi$.  The branch correction is included before transporting the unwrapped $\tilde\delta$.

The radial curvature varies as
\begin{equation}
\tilde{\calH}_{2,\rK}^{[1]}
=(\tilde{\calP}_{\K}\tilde U_1)''
-\tilde\xi_{\K}(\tilde{\calQ}_{\K}\tilde U_2)''
-\tilde{\calQ}_{\K}''\tilde\xi_{\rK}^{[1]}.
\label{eq:H21}
\end{equation}
Expanding the exact expressions in Eq.~\eqref{eq:exactfields} then gives
\begin{align}
\tilde\tau_{\rK}^{[1]}
={}&
\left[\tilde{\calP}_{\K}\tilde{\calH}_{\K}\tilde U_1
+\tilde{\calP}_{\K}\tilde{\calH}_{\rK}^{[1]}
+a\tilde\xi_{\rK}^{[1]}\right]\tilde G_{\theta,\K}
\nonumber\\
&+
\left(\tilde{\calP}_{\K}\tilde{\calH}_{\K}+a\tilde\xi_{\K}-a^2\right)
\tilde G_\theta^{[1]}
+a^2\tilde G_t^{[1]},
\label{eq:tau1r}
\end{align}
\begin{align}
\tilde\delta_{\rK}^{[1]}
={}&
\left[\tilde{\calQ}_{\K}\tilde{\calH}_{\K}\tilde U_2
+\tilde{\calQ}_{\K}\tilde{\calH}_{\rK}^{[1]}\right]\tilde G_{\theta,\K}
\nonumber\\
&+
(\tilde{\calQ}_{\K}\tilde{\calH}_{\K}-a)\tilde G_\theta^{[1]}
+\tilde\xi_{\rK}^{[1]}\tilde G_{\phi,\K}
+\tilde\xi_{\K}\tilde G_\phi^{[1]},
\label{eq:delta1r}
\end{align}
and
\begin{equation}
\frac{\tilde\gamma_{\rK}^{[1]}}{\tilde\gamma_{\K}}
=
\frac{\tilde G_\theta^{[1]}}{\tilde G_{\theta,\K}}
+\frac12\left(
\tilde U_5
+\frac{\tilde{\calH}_{\rK}^{[1]}}{\tilde{\calH}_{\K}}
+\frac{\tilde{\calH}_{2,\rK}^{[1]}}{\tilde{\calH}_{2,\K}}
\right).
\label{eq:gamma1r}
\end{equation}
These equations give the deviations directly in terms of the Kerr-deviation profiles and their radial derivatives.  They are intermediate shell-labeled expressions; the direction-resolved quantities used for comparison with concrete metrics are obtained from Eq.~\eqref{eq:fixedscreen}.  Explicitly,
\begin{subequations}
\label{eq:criticalfieldsphi}
\begin{align}
\tilde\tau^{[1]}_{\varphi}(\varphi)
&=\left[
\tilde\tau^{[1]}_{\rK}
-\frac{\dd\tilde\tau_{\K}/\dd\rK}
{\dd\tilde\varphi_{\K}/\dd\rK}
\tilde\varphi^{[1]}_{\rK}
\right]_{\rK=\tilde r_{\K}(\varphi)},
\label{eq:tau1phi}\\
\tilde\delta^{[1]}_{\varphi}(\varphi)
&=\left[
\tilde\delta^{[1]}_{\rK}
-\frac{\dd\tilde\delta_{\K}/\dd\rK}
{\dd\tilde\varphi_{\K}/\dd\rK}
\tilde\varphi^{[1]}_{\rK}
\right]_{\rK=\tilde r_{\K}(\varphi)},
\label{eq:delta1phi}\\
\tilde\gamma^{[1]}_{\varphi}(\varphi)
&=\left[
\tilde\gamma^{[1]}_{\rK}
-\frac{\dd\tilde\gamma_{\K}/\dd\rK}
{\dd\tilde\varphi_{\K}/\dd\rK}
\tilde\varphi^{[1]}_{\rK}
\right]_{\rK=\tilde r_{\K}(\varphi)}.
\label{eq:gamma1phi}
\end{align}
\end{subequations}
The profiles in Sec.~\ref{sec:results} use these common-$\varphi$ quantities.  They show directly how a chosen set of metric-deviation functions, or the dimensionless parameters defining them in a concrete metric family, changes the critical observables at each direction around the observer screen.

\subsection{First-order expansion of the intrinsic relations}
\label{subsec:weakintrinsic}

Expanding the intrinsic relations of Sec.~\ref{sec:intrinsic} identifies how $\Copt$ and radial instability determine the critical-parameter changes.  We first compare orbits at the same conserved impact parameter $\tilde\xi$, which removes the change of radial label from the optical relation.  Equation~\eqref{eq:fixedscreen} then converts the results to the common screen direction used for the observable profiles.

At fixed $\tilde\xi$, the spherical orbit shifts in the radial chart by
\begin{equation}
\tilde r_{\tilde\xi}^{[1]}
=-\frac{\tilde\xi_{\rK}^{[1]}}{\tilde\xi_{\K}'},
\label{eq:rxishift}
\end{equation}
but the Kerr spherical condition removes this label shift from the optical-shell variation.  Using Eq.~\eqref{eq:H1r} gives
\begin{equation}
\left.\tilde{\calH}^{[1]}\right|_{\tilde\xi}=\Copt,
\qquad
\left.\tilde\eta^{[1]}\right|_{\tilde\xi}
=2\tilde{\calH}_{\K}\Copt.
\label{eq:fixedxioptical}
\end{equation}
These identities provide the first-order bridge to the exact shell relations of Sec.~\ref{sec:intrinsic}.  Here and below $\Copt(\tilde\xi)$ denotes Eq.~\eqref{eq:Copt} evaluated at the Kerr shell point $\tilde r_{\K}(\tilde\xi)$.

For compactness define
\begin{equation}
\tilde B_{\tau,\K}
=\tilde\eta_{\K}-\frac{\tilde\xi}{2}\tilde\eta_{\K,\tilde\xi},
\qquad
\tilde B_{\delta,\K}
=-\tilde\xi-\frac12\tilde\eta_{\K,\tilde\xi}.
\label{eq:AtauAdelta}
\end{equation}
Expanding Eqs.~\eqref{eq:taucurve} and \eqref{eq:deltacurve} at fixed $\tilde\xi$ gives
\begin{align}
\left.\tilde\tau^{[1]}\right|_{\tilde\xi}
={}&
\left[(2\tilde{\calH}_{\K}+\tilde\xi\tilde{\calQ}_{\K})\Copt
-\tilde\xi\tilde{\calH}_{\K}\Copt_{,\tilde\xi}\right]\tilde G_{\theta,\K}
\nonumber\\
&+2\tilde{\calH}_{\K}\Copt
\left[\tilde B_{\tau,\K}(\partial_{\tilde\eta}\tilde G_\theta)_{\K}
+a^2(\partial_{\tilde\eta}\tilde G_t)_{\K}\right],
\label{eq:tau1xi}
\end{align}
\begin{align}
\left.\tilde\delta^{[1]}\right|_{\tilde\xi}
={}&
\left(\tilde{\calQ}_{\K}\Copt
-\tilde{\calH}_{\K}\Copt_{,\tilde\xi}\right)\tilde G_{\theta,\K}
\nonumber\\
&+2\tilde{\calH}_{\K}\Copt
\left[\tilde B_{\delta,\K}(\partial_{\tilde\eta}\tilde G_\theta)_{\K}
+\partial_{\tilde\eta}(\tilde\xi \tilde G_\phi)_{\K}\right],
\label{eq:delta1xi}
\end{align}
where the continuously unwrapped azimuthal convention is held fixed.  The displayed derivatives are evaluated on regular patches away from $\tilde\xi=0$; pole-crossing limits are taken on the full branch-matched expression.  At fixed $\tilde\xi$, the first-order changes of the time delay and azimuthal advance depend on the optical deviations through $\Copt(\tilde\xi)$ and its first derivative.

The Lyapunov exponent also contains the radial-instability information.  Expanding Eq.~\eqref{eq:gammaintrinsic} gives
\begin{equation}
\left.
\frac{\tilde\gamma^{[1]}}{\tilde\gamma_{\K}}
\right|_{\tilde\xi}
=
2\tilde{\calH}_{\K}\Copt
\frac{(\partial_{\tilde\eta}\tilde G_\theta)_{\K}}{\tilde G_{\theta,\K}}
+\frac12\left[
\frac{\tilde{\calV}^{[1]}}{\tilde{\calV}_{\K}}
+\frac{\Copt}{\tilde{\calH}_{\K}}
+\frac{\Copt_{,\tilde\xi\tilde\xi}}
{\tilde{\calH}_{\K,\tilde\xi\tilde\xi}}
\right].
\label{eq:gamma1xi}
\end{equation}
Write the label-matched impact-parameter change as
\begin{equation}
\tilde\ell_\xi(\rK)\equiv\tilde\xi^{[1]}_{\rK}.
\end{equation}
A fixed-$\tilde\xi$ expansion of Eq.~\eqref{eq:Vdef} gives
\begin{equation}
\left.
\frac{\tilde{\calV}^{[1]}}{\tilde{\calV}_{\K}}
\right|_{\tilde\xi}
=
\tilde U_5
+2\frac{\tilde\ell_\xi'}{\tilde\xi_{\K}'}
-\frac{\tilde\ell_\xi}{\tilde\xi_{\K}'}
\left(
\frac{\tilde{\calW}_{\K}'}{\tilde{\calW}_{\K}}
+2\frac{\tilde\xi_{\K}''}{\tilde\xi_{\K}'}
\right).
\label{eq:V1xi}
\end{equation}
For metric substitution, Eq.~\eqref{eq:Copt} gives $\Copt$ in terms of $\tilde U_1$ and $\tilde U_2$, while Eq.~\eqref{eq:labelmatchedvariations} supplies $\tilde\ell_\xi$ and hence its radial derivative.  Substitution into Eq.~\eqref{eq:V1xi} then gives the additional radial-instability variation in terms of $(\tilde U_1,\tilde U_2,\tilde U_5)$.  The prime on $\tilde\ell_\xi$ in Eq.~\eqref{eq:V1xi} is a radial derivative along the reference shell.  Derivatives with respect to $\tilde\xi$ can be evaluated using $\partial_{\tilde\xi}=(\tilde\xi_{\K}')^{-1}\dd/\dd\rK$.  The time-delay and azimuthal changes require $\Copt$ and its first derivative, while the instability exponent additionally involves its second derivative and $\tilde{\calV}^{[1]}$.  Equations~\eqref{eq:tau1r}--\eqref{eq:criticalfieldsphi} are convenient for direct metric substitution; Eqs.~\eqref{eq:fixedxioptical}--\eqref{eq:V1xi} expose the relations between the resulting changes.

The derivative order has a physical interpretation.  The curve position measures the optical perturbation itself.  The time and azimuth accumulated along an orbit also depend on how the optical shell varies between neighboring impact parameters.  Radial instability probes the curvature of this optical relation as well as its kinetic normalization.  Accordingly, knowledge of $\Copt$ at one orbit does not supply its first or second derivatives, even when the local normal displacement is zero.

A functional optical null direction $\Copt\equiv0$ leaves the complete first-order critical curve, $\tilde\tau$, and $\tilde\delta$ unchanged, while
\begin{equation}
\Copt\equiv0
\quad\Longrightarrow\quad
\left.
\frac{\tilde\gamma^{[1]}}{\tilde\gamma_{\K}}
\right|_{\tilde\xi}
=\frac12
\left.
\frac{\tilde{\calV}^{[1]}}{\tilde{\calV}_{\K}}
\right|_{\tilde\xi}.
\label{eq:gammanull}
\end{equation}
For a genuine radial coordinate transformation $\tilde{\calV}^{[1]}$ also vanishes, so all critical observables are unchanged.  The proof is given in Appendix~\ref{app:relabelpolar}.

\section{Metric sectors and representative examples}
\label{sec:results}

\begin{figure*}[t]
\centering
\includegraphics[width=0.91\textwidth]{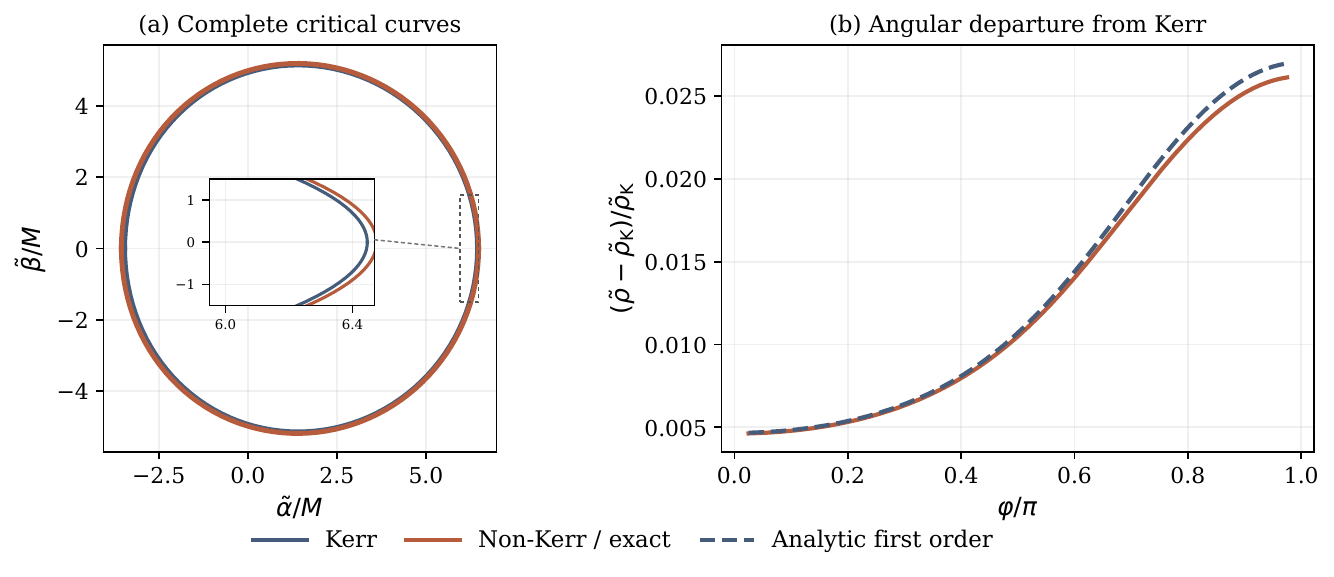}
\caption{Geometric interpretation of a Kerr-relative critical-curve deviation for an asymptotic observer at $a/M=0.8$ and $\theta_o=60^\circ$.  (a) The complete upper and lower critical-curve branches for Kerr and a controlled optical deformation $a_1=1+\chi(M/r)^3$, with $a_2=a_5=1$ and $\chi=0.20$; the inset magnifies the right-hand edge marked by a thin box on the full curve, using the same actual curves without rescaling their relative displacement.  (b) The corresponding fractional radial difference as a function of the observer-screen angle.  Solid curves are obtained from the exact finite-deformation metric, while the dashed curve is calculated directly from the analytic first-order formula.  The example illustrates how a small geometric displacement becomes a discernible angular profile.}
\label{fig:curvegeometry}
\end{figure*}

The examples show how optical and radial kinetic deformations change the critical curve and the three critical-parameter profiles.  Figure~\ref{fig:curvegeometry} first relates the complete curve to its angular departure from Kerr.  We then vary the radial functions separately to identify their contributions, before applying the same formulas to concrete metrics.  A comparison at matched polar critical-circle radius illustrates the additional radial information sampled at finite inclination.  Detailed mappings of representative metrics to the Kerr-relative functions $(h,f,a_1,a_2,a_5)$ are given in Table~II of Ref.~\cite{YanEtAl2026Polar}.  Unless stated otherwise, $M=1$.

\subsection{Optical and radial kinetic deformations}
\label{subsec:channelsphysical}

\begin{figure*}[t]
\centering
\includegraphics[width=0.985\textwidth]{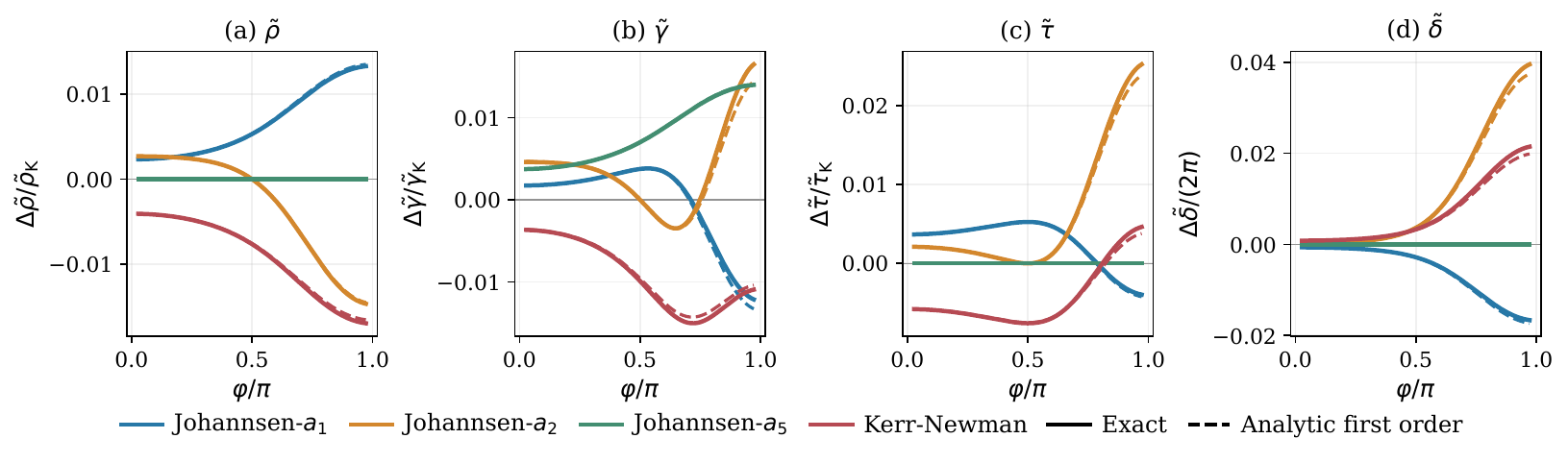}
\caption{Kerr-relative departures of the critical-curve radius and three photon-ring critical parameters at $M=1$, $a/M=0.8$, $\theta_o=60^\circ$, and family-relative amplitude $\hat\chi=0.10$.  Solid curves are exact finite-family differences and dashed curves use the analytic first-order expressions.  Panels (a)--(d) show $\tilde\rho$, $\tilde\gamma$, $\tilde\tau$, and $\tilde\delta$, respectively.  The controlled $a_1$, $a_2$, and $a_5$ families isolate optical or radial kinetic ingredients, while Kerr--Newman supplies a physical $\Delta$-sector example.  Positive $a_1$ gives outward normal motion, the pure $a_2$ shape change has a node at $\tilde\alpha=0$, and $a_5$ changes only the instability exponent.  The family-dependent amplitudes are plotting conventions and do not define a model-independent sensitivity ranking.}
\label{fig:channels}
\end{figure*}

The functions $a_1$ and $a_2$ modify the optical combinations $\calP$ and $\calQ$, whereas $a_5$ changes the radial kinetic factor $\calW$.  To display these directions separately, Figure~\ref{fig:channels} uses the one-parameter profiles $a_1=1+\chi(M/r)^3$, $a_2=1+\chi(M/r)^2$, and $a_5=1+\chi(M/r)^2$, switching on one function at a time; a Kerr--Newman $\Delta$ deformation is included for comparison.  The exact finite-family deviations are compared with first-order predictions calculated directly from the analytic expressions of Sec.~\ref{sec:weak}.  We use the family-relative plotting coordinate $\hat\chi=\chi/\chi_\star$, with $\chi_\star=1$ for the controlled profiles and $\chi_\star^{\rm KN}=1-(a/M)^2$ for Kerr--Newman.  This is only a plotting convention, not a model-independent norm of non-Kerr deformation.

For a positive first-order $a_1$ profile, the optical combination is $\Copt=\tilde{\calP}_{\K}\tilde U_1>0$ on the regular exterior shell.  Since $\tilde F_{\K}>0$ in Eq.~\eqref{eq:Kerrnormalfactor}, the first-order normal displacement is outward throughout the visible curve.  The magnitude varies with the Kerr orbit and observer inclination, and need not represent a uniform rescaling.  For the profile $U_1=(M/r)^3$ in Figure~\ref{fig:channels}, the radial weight grows inward, with the screen dependence further determined by $\tilde F_{\K}\tilde{\calP}_{\K}$.

For the pure $a_2$ deformation,
$\Copt=-\tilde\xi_{\K}\tilde{\calQ}_{\K}\tilde U_2$ changes sign when the shell crosses
$\tilde\xi_{\K}=0$, giving a stationary normal-shape node at
$\tilde\alpha=0$.  In fact this screen point remains fixed for the
finite pure-$a_2$ family: at $\tilde\xi=0$ the exact optical function
is $\tilde{\calP}$, unchanged from Kerr, and its stationary value is attained at
the same $\tilde{\calP}'=0$ orbit.  Consequently the associated screen ordinate
is unchanged as well.

The pure kinetic $a_5$ deformation leaves the complete critical curve, $\tilde\tau$, and $\tilde\delta$ unchanged while changing the instability exponent.  At first order,
\begin{equation}
\tilde\rho^{[1]}=\tilde\tau^{[1]}=\tilde\delta^{[1]}=0,
\qquad
\frac{\tilde\gamma^{[1]}}{\tilde\gamma_{\K}}=\frac12\tilde U_5.
\label{eq:pureU5}
\end{equation}
For the pure kinetic family the exact relation is $\tilde\gamma/\tilde\gamma_{\K}=\sqrt{1+\chi(M/\tilde r)^2}$, while the other three profiles remain Kerr.  This gives an exact check of the radial-normalization contribution.

The rotating Simpson--Visser geometry realizes this purely kinetic mechanism \cite{SimpsonVisser2019,MazzaFranzinLiberati2021,Shaikh2021}.  In the radial representation used here and in Ref.~\cite{SalehiBroderickGeorgiev2024}, its length parameter $\ell_{\rm SV}$ enters through
\begin{equation}
\calP=\calP_{\K},\quad
\calQ=\calQ_{\K},\quad
\calW=\calW_{\K}\left(1-\frac{\ell_{\rm SV}^2}{r^2}\right),
\label{eq:SVPQW}
\end{equation}
so the complete critical curve remains Kerr at arbitrary inclination.  At a common screen origin and position angle,
\begin{equation}
\tilde\rho_{\rm SV}=\tilde\rho_{\K},\qquad
\tilde\tau_{\rm SV}=\tilde\tau_{\K},\qquad
\tilde\delta_{\rm SV}=\tilde\delta_{\K},
\label{eq:SVblind}
\end{equation}
while
\begin{equation}
\frac{\tilde\gamma_{\rm SV}(\tilde r)}{\tilde\gamma_{\K}(\tilde r)}
=\sqrt{1-\frac{\ell_{\rm SV}^2}{\tilde r^{\,2}}}.
\label{eq:SVgamma}
\end{equation}
These exact relations apply when the same exterior unstable shell remains accessible, with $\tilde r>\ell_{\rm SV}$; throat orbits require a separate global analysis.  The small-$\ell_{\rm SV}^2/M^2$ expansion of Eq.~\eqref{eq:SVgamma} agrees with Eq.~\eqref{eq:pureU5}.

Kerr--Newman supplies a pure $\Delta$ example: $\Delta=\Delta_{\K}+q^2$, $f=0$, and $a_1=a_2=a_5=1$ \cite{NewmanEtAl1965}.  Here $q$ is the electric charge and $\chi=q^2/M^2$.  The deformation changes both optical functions and the radial kinetic factor, with correlated first-order profiles given by Eq.~\eqref{eq:lowerUmap}.  Positive $q^2$ produces an inward first-order normal displacement, as follows from the general $\Delta$-sector law in Eq.~\eqref{eq:DeltaNormalSign} below.

Kerr--Sen is a mixed $\Delta$-and-$a_1$ example in the radial chart used here: $\Delta=r(r+\sigma)-2Mr+a^2$, $f=\sigma r$, $a_1=1+\sigma r/(r^2+a^2)$, and $a_2=a_5=1$~\cite{Sen1992,YanEtAl2026Polar}.  The parameter $\sigma$ has dimensions of length, and we use $\chi=\sigma/M$.  Its contributions to the leading asymptotic coefficients of the full metric compensate, leaving the ADM mass $M$ and angular momentum $J=Ma$ fixed.  Thus a change of $\Delta$ alone would not represent this family. Analytic studies of near-extremal Kerr--Sen also describe near-horizon emitter images \cite{GuoSongYan2020}.  With these definitions of $\chi$, the difference between the Kerr--Newman and Kerr--Sen first-order profiles is a radial-relabeling term.  Appendix~\ref{app:relabelpolar} checks its cancellation from the critical observables, including the radial kinetic contribution.

The optical deformations in Figures~\ref{fig:curvegeometry} and \ref{fig:channels}, and the $\Delta$ profiles of Figure~\ref{fig:sampling} below, show enhanced departures toward large $\varphi$ on the upper branch.  With $a>0$, increasing $\varphi$ moves toward $\tilde\alpha<0$ and $\tilde\xi>0$, sampling smaller, prograde orbital radii.  The displayed radial profiles strengthen inward, with their screen dependence further weighted by the Kerr geometrical factors.  Finite-amplitude corrections also tend to appear first on this side in these examples.

\subsection{The \texorpdfstring{$\Delta$}{Delta} class and representative examples}
\label{subsec:Deltaclass}

A particularly broad physical subclass is obtained by keeping $f=0$ and $a_1=a_2=a_5=1$ while replacing the Kerr radial function by a generalized $\Delta(r)$.  Then
\begin{equation}
\calP=\frac{r^2+a^2}{\sqrt{\Delta}},
\qquad
\calQ=\frac{a}{\sqrt{\Delta}},
\qquad
\calW=\Delta.
\label{eq:DeltaclassExact}
\end{equation}
This includes Kerr--Newman \cite{NewmanEtAl1965} and the tidal-charge geometry \cite{AlievGumrukcuoglu2005}, as well as Carter-separable rotating mass-function realizations used for Bardeen-like and Hayward-like black holes \cite{BambiModesto2013,AzregAinou2014}.  For $\Delta=r^2-2rm(r)+a^2$, the critical impact parameters below reproduce the general shadow formulas of Tsukamoto, Li, and Bambi and Tsukamoto \cite{TsukamotoLiBambi2014,Tsukamoto2018}.  Analytic photon-region constructions in other separable families also use a double radial root \cite{GrenzebachEtAl2014}.  Here the Kerr angular structure is retained, so the same angular integrals give the three critical parameters throughout the class.  With $\tilde\Delta^{(n)}=\Delta^{(n)}(\tilde r)$,
\begin{equation}
\tilde\xi
=\frac{(\tilde r^2+a^2)\tilde\Delta'-4\tilde r\tilde\Delta}{a\tilde\Delta'},
\label{eq:Deltaxi}
\end{equation}
\begin{equation}
\tilde{\calH}=\frac{4\tilde r\sqrt{\tilde\Delta}}{\tilde\Delta'},
\qquad
\tilde\eta=\frac{16\tilde r^2\tilde\Delta}{(\tilde\Delta')^2}-(\tilde\xi-a)^2,
\label{eq:DeltaetaH}
\end{equation}
and
\begin{equation}
\tilde{\calH}_2
=\frac{1}{\sqrt{\tilde\Delta}}
\left(2+\frac{\tilde r\tilde\Delta'}{\tilde\Delta}-\frac{2\tilde r\tilde\Delta''}{\tilde\Delta'}\right).
\label{eq:DeltaH2}
\end{equation}
The exact critical parameters follow as
\begin{equation}
\tilde\gamma
=\tilde G_\theta
\sqrt{\frac{4\tilde r\tilde\Delta}{\tilde\Delta'}
\left(2+\frac{\tilde r\tilde\Delta'}{\tilde\Delta}-\frac{2\tilde r\tilde\Delta''}{\tilde\Delta'}\right)},
\label{eq:Deltagamma}
\end{equation}
\begin{equation}
\tilde\tau
=\frac{\tilde r\left[4(\tilde r^2+a^2-\tilde\Delta)+\tilde r\tilde\Delta'\right]}{\tilde\Delta'}\tilde G_\theta+a^2\tilde G_t,
\label{eq:Deltatau}
\end{equation}
and
\begin{equation}
\tilde\delta
=\frac{a(4\tilde r-\tilde\Delta')}{\tilde\Delta'}\tilde G_\theta+\tilde\xi \tilde G_\phi
+2\pi\Heav(\tilde r-\tilde r_0).
\label{eq:Deltadelta}
\end{equation}
With matched Carter-constant conventions, Eqs.~\eqref{eq:Deltagamma}--\eqref{eq:Deltadelta} recover the critical parameters studied for explicit non-Kerr metrics in Ref.~\cite{WaliaKocherlakotaChangSalehi2025}.

The weak-deviation limit gives a direct normal-deformation law.  For a weak deviation $\Delta=\Delta_{\K}(1+\varepsilon\epsilon_\Delta)$,
\begin{equation}
U_1=U_2=-\frac12\epsilon_\Delta,
\qquad U_5=\epsilon_\Delta,
\end{equation}
so the first-order normal deformation is controlled by
\begin{equation}
\Copt_\Delta=-\frac12\tilde\epsilon_\Delta\,\tilde{\calH}_{\K}.
\label{eq:CDelta}
\end{equation}

Writing $\Delta=\Delta_{\K}+\varepsilon h_1(r)$, Eq.~\eqref{eq:CDelta}
becomes the direct normal-deformation law
\begin{equation}
\Copt_{\Delta}=-\frac{\tilde{\calH}_{\K}}{2\tilde\Delta_{\K}}\tilde h_1,
\qquad
\tilde u_{n,\Delta}=-\frac{\tilde F_{\K}\tilde{\calH}_{\K}}{2\tilde\Delta_{\K}}\tilde h_1.
\label{eq:DeltaNormalSign}
\end{equation}
On the regular exterior shell, where $\tilde F_{\K}$, $\tilde\Delta_{\K}$, and
$\tilde{\calH}_{\K}$ are positive, $\tilde h_1>0$ implies inward first-order
normal motion, while $\tilde h_1<0$ implies outward motion.  If the sign is
uniform over the visible shell, the same sign holds for the first-order
projected support positions and their corresponding diameters, provided
the support points remain regular.  This local first-order statement does
not establish global inclusion of finite-deformation shadows.

\begin{figure*}[t!]
\centering
\includegraphics[width=0.985\textwidth]{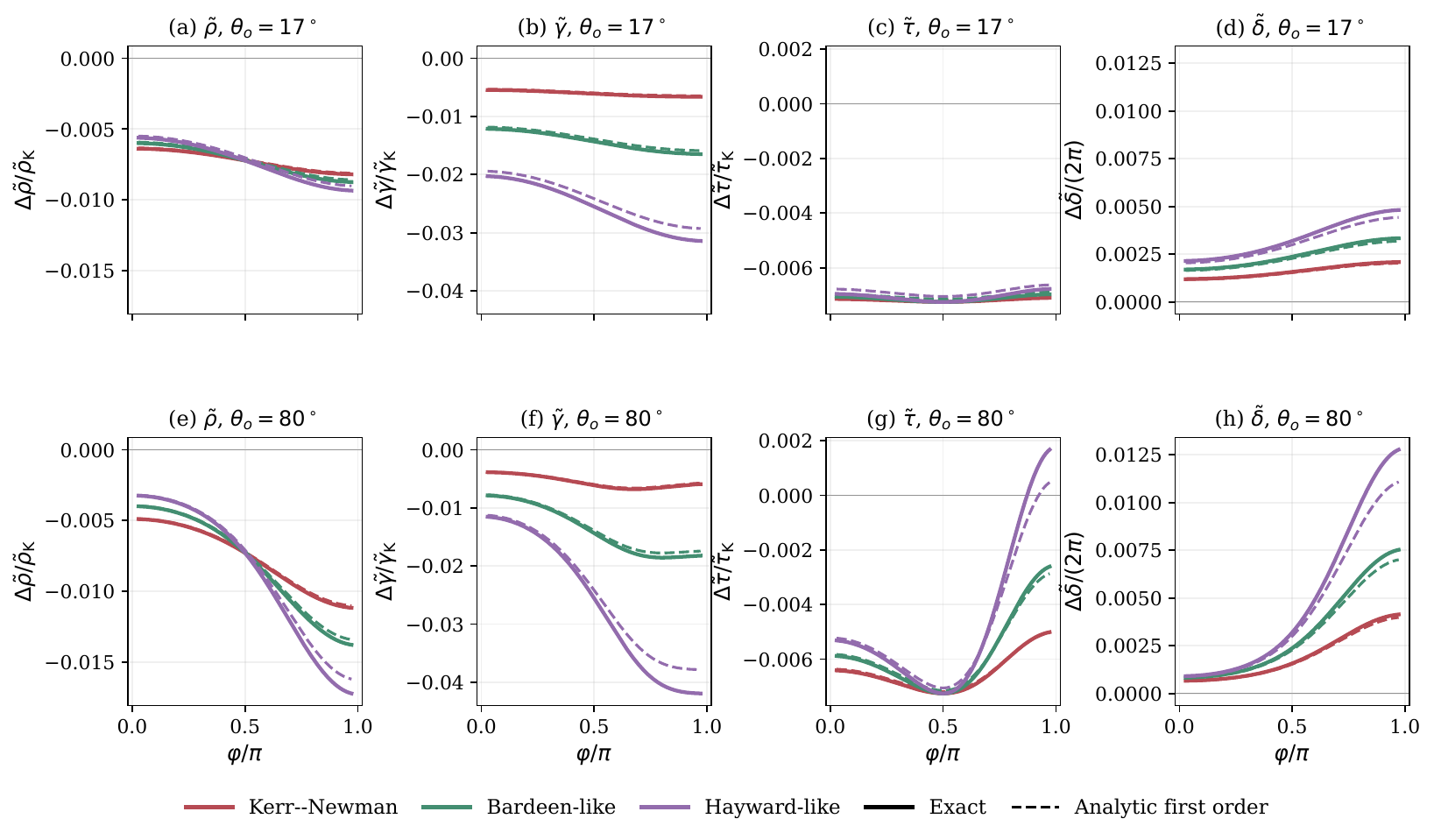}
\caption{Three representative $\Delta(r)$ geometries matched to the same exact polar critical radius at $a/M=0.5$: Kerr--Newman, and the Carter-separable Bardeen-like and Hayward-like realizations.  Upper and lower rows show observer inclinations $\theta_o=17^\circ$ and $80^\circ$, respectively; corresponding columns share the same vertical scale, so inclination-dependent differences can be compared directly.  Solid curves show exact finite-family departures from Kerr, while dashed curves come from the independent analytic first-order formulas.  The polar-radius matching fixes a single datum, whereas finite inclination samples the radial profiles across a nonzero visible-shell interval.  The exact matching at finite parameter values need not be reproduced identically by a first-order truncation.}
\label{fig:sampling}
\end{figure*}

Several physical examples share a constant radial correction.  At fixed asymptotic mass and spin,
\begin{equation}
\Delta=\Delta_{\K}+c
\label{eq:constantc}
\end{equation}
has the same vacuum null-geodesic problem for a common value of $c$.  Kerr--Newman corresponds to $c=q^2$, the usual rotating braneworld metric to $c=\beta$, and Kerr--MOG to $c=\alpha M^2/(1+\alpha)$ when $M$ denotes the ADM mass \cite{SheoranHerreraNucamendi2018}.  The physical interpretation and allowed parameter range differ between the theories---in particular the tidal charge may be negative---but the critical observables considered here are governed by one exact radial family.  For this subclass,
\begin{equation}
\tilde\xi_c
=\frac{\tilde r^3-3M\tilde r^2+a^2(\tilde r+M)+2c\tilde r}{a(M-\tilde r)},
\label{eq:constantcxi}
\end{equation}
\begin{equation}
\tilde\eta_c
=\frac{\tilde r^2\left[4a^2(M\tilde r-c)-(\tilde r^2-3M\tilde r+2c)^2\right]}
{a^2(\tilde r-M)^2},
\label{eq:constantceta}
\end{equation}
These expressions recover the critical-impact-parameter formulas used for the braneworld and Kerr--MOG cases \cite{AmarillaEiroa2012,GuoObersYan2018,SheoranHerreraNucamendi2018}.  Their shadow trends correspond to different parameter slices of this exact radial family.

Different radial dependences arise in the Carter-separable mass-function realizations
\begin{equation}
m_B(r)=\frac{Mr^3}{(r^2+g^2)^{3/2}},
\qquad
m_H(r)=\frac{Mr^3}{r^3+2M\ell^2},
\label{eq:BHmassfunctions}
\end{equation}
inserted through $\Delta=r^2-2rm(r)+a^2$~\cite{WaliaKocherlakotaChangSalehi2025}.  The Hayward mass function originates in a nonsingular spherical model \cite{Hayward2006}; the rotating realizations used here are the specified Kerr-like constructions \cite{BambiModesto2013,AzregAinou2014,AbdujabbarovEtAl2016}, rather than a uniqueness claim for rotating regular black holes.  The comparison concerns their exterior null dynamics.  To test how much a polar-size match constrains finite-inclination profiles, we compare these metrics with Kerr--Newman at $a/M=0.5$, choosing the same exact polar critical-curve radius,
\begin{equation}
\tilde b_p=5.083575864\,M,
\end{equation}
with
\begin{equation}
\begin{aligned}
\frac{q^2}{M^2}&=0.04,\\
\frac{g^2}{M^2}&=0.03822134,\\
\frac{\ell^2}{M^2}&=0.08156634.
\end{aligned}
\label{eq:matchedparams}
\end{equation}
The parameters match the polar radius in the exact finite-deformation metrics.  A first-order match instead requires equal $\Copt$ at the Kerr pole-crossing orbit, so the linearized radii need not coincide for these parameter values.  Equation~\eqref{eq:polartau} also fixes the exact polar time delay at the common radius and specified $a$, while the polar azimuthal advance and instability can differ.

All profiles use a common screen origin, so changes of $\tilde\rho$ include position, size, and noncircular shape.  For the examples in Figure~\ref{fig:sampling}, the matched metrics have closely spaced critical-curve radius and time-delay profiles at low inclination, while their differences become more pronounced at higher inclination.  Their demagnification profiles are already distinct at low inclination, with a separation that changes less strongly with inclination.

As inclination increases, the broader visible shell interval samples radial variations in $\Delta$ and its derivatives beyond those fixed by a polar-radius match.  The intrinsic relations of Sec.~\ref{sec:intrinsic} explain the different behavior of the critical parameters: the instability already probes radial curvature at the polar orbit, while the finite-inclination time-delay and azimuthal profiles also sample the optical relation over neighboring orbits.

\section{Discussion and conclusions}
\label{sec:discussion}

We have derived analytic first-order changes of the complete critical curve and photon-ring critical parameters for general observer inclinations in the separable metric class defined in Sec.~\ref{sec:metric}.  The Kerr-relative metric decomposition makes the optical and radial kinetic contributions explicit.  The signed normal displacement factorizes as $\tilde u_n=\tilde F_{\K}\Copt$, with all optical metric deviations entering through $\Copt$.  Orbit-label shifts contribute tangentially and cancel from this projection.  The accompanying critical-parameter formulas give the demagnification exponent, time delay, and azimuthal advance at the same observer-screen direction as the curve radius.

The intrinsic relations among the exact observables explain these dependencies.  At specified global parameters, a differentiable critical locus determines the ideal time-delay and azimuthal-advance profiles, while the demagnification exponent additionally requires the radial-instability function $\tilde{\calV}$.  Their first-order expansion shows that the curve displacement depends on $\Copt$, time delay and azimuthal advance also depend on its first derivative, and instability involves its second derivative together with the change of $\tilde{\calV}$.  The distinction separates optical changes from changes in the radial escape of nearby rays.

The metric examples illustrate the physical content of the formulas.  A positive $h_1$ in the $\Delta$ class produces inward normal motion, a pure $a_2$ deformation preserves the pole-crossing screen point, and a radial kinetic deformation changes instability while leaving the curve, time delay, and azimuthal advance fixed.  Metrics with matched polar critical-circle radius develop different angular profiles as finite inclination samples a wider radial interval.  Comparisons between exact and linear profiles identify the finite-amplitude corrections, supported by independent numerical checks.

Imaging, timing, and azimuthal measurements access these related geometric quantities through different source and instrumental effects \cite{GrallaObservable2020,Hadar2021,SalehiWaliaChangKocherlakota2025}.  The analytic formulas can supply geometric inputs to compressed shape models or visibility-domain analyses \cite{MedeirosEtAl2020,BHEXScience2024}.  An observational application must additionally account for finite-order emission, ring width, variability, scattering, and measurement covariance \cite{PaugnatEtAl2022,CardenasLupsasca2023,SalehiBroderick2026}, with ray tracing and radiative transfer connecting the geometry to an emitting source \cite{YounsiEtAl2016,CardenasEtAl2023AART}. Kerr thick-disk and jet calculations and non-Kerr thin-disk images illustrate the effects of emission anisotropy, flow motion, and inclination \cite{ZhangHouGuoChen2024,HeLiuYangZeng2026}.  Comparisons of instability measures also require a common convention for the orbital interval \cite{DeichEtAl2024}.

The formulas apply to regular, accessible exterior shell branches with Kerr angular structure; the conformal function $f(r)$ cancels from the vacuum null observables considered here.  Angular deformations, nonseparable dynamics, and degenerate shell limits require separate treatments \cite{CunhaHerdeiro2018,PerlickTsupko2022,GlampedakisPappas2019}.  Using the intrinsic relations with measured curves also requires control of errors in their derivatives.  Within this domain, the formulas give the first-order changes of the critical observables at each direction on the observer screen for a specified radial metric deformation.
\begin{acknowledgments}
This work was supported by the National Natural Science Foundation of China under Grant No.~12305070, and the Basic Research Program of Shanxi Province under Grant Nos.~202303021222018 and 202303021221033.
\end{acknowledgments}

\appendix

\section{Radial relabeling and strict-polar consistency}
\label{app:relabelpolar}
\label{app:proofs}
\label{app:polar}

This appendix establishes the radial-relabeling statements used in Sec.~\ref{sec:weak} and collects the strict-polar relations used in Sec.~\ref{sec:results}.

\subsection{Radial relabeling}

Consider an infinitesimal radial reparameterization
\begin{equation}
R=r+\varepsilon s(r).
\end{equation}
The Kerr optical functions represented in the new label obey
\begin{equation}
\calP(r)=\calP_{\K}(R)
=\calP_{\K}(r)+\varepsilon s\calP_{\K}'+\ord(\varepsilon^2),
\end{equation}
\begin{equation}
\calQ(r)=\calQ_{\K}(R)
=\calQ_{\K}(r)+\varepsilon s\calQ_{\K}'+\ord(\varepsilon^2).
\end{equation}
Thus Eq.~\eqref{eq:nullmap} is precisely the optical tangent generated by a radial relabeling.  Conversely, on a regular Kerr shell patch with $\tilde{\calQ}_{\K}'\neq0$, $\Copt=0$ implies $\tilde{\calP}^{[1]}=\tilde\xi_{\K}\tilde{\calQ}^{[1]}$.  Using $\tilde{\calP}_{\K}'=\tilde\xi_{\K}\tilde{\calQ}_{\K}'$ and setting $\tilde s=\tilde{\calQ}^{[1]}/\tilde{\calQ}_{\K}'$ gives Eq.~\eqref{eq:nullmap}.  Substitution into Eq.~\eqref{eq:labelmatchedvariations} yields
\begin{equation}
\tilde\xi_{\rK}^{[1]}=\tilde s\tilde\xi_{\K}',
\qquad
\tilde{\vect x}_{\rK}^{[1]}=\tilde s\frac{\dd\tilde{\vect x}_{\K}}{\dd\rK},
\label{eq:tangent}
\end{equation}
so the displacement of the screen curve is purely tangential.

The radial equation transforms as
\begin{equation}
\left(\frac{\dd r}{\dd\nu}\right)^2
=\frac{\calW_{\K}(R)}{(1+\varepsilon s')^2}
[\calH^2-\zeta^2],
\end{equation}
which gives
\begin{equation}
U_5^{(g)}
=s\frac{\calW_{\K}'}{\calW_{\K}}-2s'.
\label{eq:U5gauge}
\end{equation}
Substituting $\tilde\xi_{\rK}^{[1]}=\tilde s\tilde\xi_{\K}'$ and Eq.~\eqref{eq:U5gauge} into Eq.~\eqref{eq:V1xi} gives $\tilde{\calV}^{[1]}=0$ identically.  A full radial coordinate change therefore leaves the complete critical curve of Sec.~\ref{subsec:shellcurve} and the three parameters $(\tilde\gamma,\tilde\tau,\tilde\delta)$ invariant.  If only the optical functions satisfy the relabeling condition while $\calW$ does not transform according to Eq.~\eqref{eq:U5gauge}, the critical curve, $\tilde\tau$, and $\tilde\delta$ remain unchanged but $\tilde\gamma$ can still probe the residual change of $\tilde{\calV}$.

Using the fixed-$(M,a)$ Kerr--Newman and Kerr--Sen parametrizations of Sec.~\ref{subsec:channelsphysical}, the first-order profiles per unit $\chi$ differ by
\begin{equation}
\left(U_1,U_2,U_5\right)_{\rm KS}
-\left(U_1,U_2,U_5\right)_{\rm KN}
=
\frac{M}{2}
\left(
\frac{\calP_{\K}'}{\calP_{\K}},
\frac{\calQ_{\K}'}{\calQ_{\K}},
\frac{\calW_{\K}'}{\calW_{\K}}
\right),
\label{eq:KNKSgauge}
\end{equation}
The right-hand side is the radial-coordinate tangent generated by $s=M/2$.  Including the radial kinetic transformation in Eq.~\eqref{eq:U5gauge} therefore removes this term from the first-order critical observables under the stated parameter matching.  Finite-parameter differences remain \cite{XavierCunhaCrispinoHerdeiro2020}.

\subsection{Strict-polar consistency}

A strict polar observer selects the pole-crossing orbit $\tilde r_p=\tilde r_0$ of Eq.~\eqref{eq:deltaexact}, characterized by
\begin{equation}
\tilde\xi(\tilde r_p)=0,
\qquad
\tilde{\calP}'(\tilde r_p)=0.
\label{eq:polarcondition}
\end{equation}
Write $\tilde b_p=\tilde{\calP}(\tilde r_p)$, $\tilde\eta_p=\tilde b_p^2-a^2>0$, and $\tilde m_p=-a^2/\tilde\eta_p$.  Substitution of $\tilde u_+=1$ and $\tilde u_-=-\tilde\eta_p/a^2$ into Eqs.~\eqref{eq:ellipticG} and \eqref{eq:exactfields} gives
\begin{equation}
\tilde\tau_p=2\sqrt{\tilde\eta_p}\,\mathrm E(\tilde m_p).
\label{eq:polartau}
\end{equation}
Thus the polar radius fixes the polar time delay at specified $a$.  In the continuous azimuthal convention of Eq.~\eqref{eq:deltaexact},
\begin{equation}
\tilde\delta_p=\pi+
(\tilde{\calQ}_p\tilde b_p-a)\tilde G_{\theta,p},
\label{eq:polardelta}
\end{equation}
which also samples $\tilde{\calQ}_p$ and is not fixed by $\tilde b_p$ alone.
At the Kerr pole-crossing orbit, $\tilde b_{\K}=\tilde{\calP}_{\K}(\tilde r_{p,\K})$ and
\begin{equation}
\Copt(\tilde r_{p,\K})
=\tilde b_{\K}\tilde U_1(\tilde r_{p,\K}).
\end{equation}
Thus the single condition $\Copt(\tilde r_{p,\K})=0$ fixes $\tilde U_1$ at that orbit and leaves the polar time delay unchanged at first order.  It does not impose the functional relation in Eq.~\eqref{eq:nullmap} on a neighborhood or constrain $\tilde U_2$ and $\tilde U_1'$, so azimuthal advance and instability can still change.  The explicit strict-polar formulas agree with the companion polar analysis~\cite{YanEtAl2026Polar}.

For a genuine radial relabeling,
\begin{equation}
U_1=s\frac{\calP_{\K}'}{\calP_{\K}},
\qquad
U_2=s\frac{\calQ_{\K}'}{\calQ_{\K}},
\qquad
U_5=s\frac{\calW_{\K}'}{\calW_{\K}}-2s'.
\label{eq:gaugeU125}
\end{equation}
At the polar orbit $\tilde{\calP}_{\K}'=0$, and substitution of Eq.~\eqref{eq:gaugeU125} into the strict-polar expressions gives zero for the critical-radius, time-delay, azimuthal-advance, and Lyapunov corrections, as required for a coordinate transformation.  The distinction is therefore between a local condition at the single orbit selected by a polar observer and a functional relabeling of a finite shell neighborhood.

% Balance the final appendix columns without changing the earlier pages.
\makeatletter
\@booleanfalse\raggedcolumn@sw
\makeatother
\flushbottom
\predisplaypenalty=0
\section{Projected positions and diameters from normal deformation}
\label{app:projection}

For a smooth convex critical curve on the observer screen, consider the supporting projected position in the direction
$\vect n_{\psi}=(\cos\psi,\sin\psi)$,
\begin{equation}
\tilde p_{\varepsilon}(\psi)=\max_{s}\bigl[\vect n_{\psi}\cdot\tilde{\vect x}_{\varepsilon}(s)\bigr],
\label{eq:supportdefinition}
\end{equation}
where $s$ parameterizes the reference Kerr curve and the perturbed curve is compared using any smooth local correspondence.  Suppose the Kerr support point $s_{\psi,\K}$ is unique and nondegenerate.  Its defining condition is
\begin{equation}
\vect n_{\psi}\cdot\tilde{\vect x}_{\K}'(s_{\psi,\K})=0.
\end{equation}
Writing $\tilde{\vect x}_{\varepsilon}(s)=\tilde{\vect x}_{\K}(s)+\varepsilon\,\delta\tilde{\vect x}(s)+\mathcal O(\varepsilon^2)$ and $s_{\psi}(\varepsilon)=s_{\psi,\K}+\varepsilon s_{\psi}^{[1]}+\mathcal O(\varepsilon^2)$ gives
\begin{align}
\tilde p^{[1]}(\psi)
&=\vect n_{\psi}\cdot\bigl[\delta\tilde{\vect x}
+s_{\psi}^{[1]}\tilde{\vect x}_{\K}'\bigr]_{s=s_{\psi,\K}}\nonumber\\
&=\vect n_{\psi}\cdot\delta\tilde{\vect x}(s_{\psi,\K}).
\label{eq:supportstationary}
\end{align}
At a regular convex support point the outward unit normal is $\vect n_{\psi}$.  The last expression is therefore the signed normal displacement $\tilde u_n(s_{\psi,\K})$; tangential relabeling does not affect it.  The projected diameter is the full width in that direction,
\begin{equation}
\tilde d_{\varepsilon}(\psi)=\tilde p_{\varepsilon}(\psi)+\tilde p_{\varepsilon}(\psi+\pi),
\label{eq:projecteddiameter}
\end{equation}
so the two projected variations are
\begin{equation}
\begin{aligned}
\tilde p^{[1]}(\psi)&=\tilde u_n(s_{\psi,\K}),\\
\tilde d^{[1]}(\psi)&=\tilde u_n(s_{\psi,\K})+\tilde u_n(s_{\psi+\pi,\K}).
\end{aligned}
\label{eq:projecteddeviations}
\end{equation}
Equation~\eqref{eq:un} supplies these shifts by evaluating $\tilde F_{\K}\Copt$ at the Kerr support points.  The support-normal direction $\psi$ need not equal the radial screen angle $\varphi$.  Relating these ideal geometric diameters to a finite-order observed ring requires an emission and transfer model.

\end{document}